# Ab-initio insights into the mechanical, phonon, bonding, electronic, optical and thermal properties of hexagonal $W_2N_3$ for potential applications

Istiak Ahmed, F. Parvin, R. S. Islam, S. H. Naqib*

Department of Physics, University of Rajshahi, Rajshahi 6205, Bangladesh

*Corresponding author; Email: salehnaqib@yahoo.com

**Abstract**

We thoroughly investigated the structural, mechanical, electronic, vibrational, optical, thermodynamic and a number of thermophysical properties of $W_2N_3$ in this study through first-principles calculations utilizing the DFT based formalisms**.** The calculated lattice parameters show excellent agreement with the available theoretical and experimental results. The mechanical and dynamical stabilities of this compound have been confirmed theoretically from the calculated elastic constants and phonon dispersion curves of $W_2N_3$. The Pugh's and Poisson's ratios of $W_2N_3$ are located quite close to the brittle/ductile borderline. $W_2N_3$ is elastically anisotropic. The calculated electronic band structure and density of states reveal that $W_2N_3$ is metallic in nature. The Fermi surface topology has also been explored. The analysis of charge density distribution map clearly shows that W atoms have comparatively high electron density around than the N atoms. Presence of covalent bondings between W-N, W-W, and N-N atoms are anticipated. High melting temperature and high phonon thermal conductivity at room temperature of $W_2N_3$ imply that the compound has potential to be used as a heat sink system. The optical characteristics demonstrate anisotropy for $W_2N_3$. The compound can be used in optoelectronic device applications due to its high absorption coefficient and low reflectivity in the visible to ultraviolet spectrum. Furthermore, the quasi-harmonic Debye model is used to examine temperature and pressure dependent thermal characteristics of $W_2N_3$ for the first time.

## 1. Introduction

Due to the importance in both fundamental science and technological applications, the investigation of superhard materials has always been an interesting topic [1-3]. Researchers have synthesized two groups of superhard materials; one of them is formed by the light elements (e.g. B, C, N, O etc.) such as diamond, cubic boron nitride and carbon nitrides [4-7], the other group comprises the borides, carbides, nitrides and oxides of transition metals (TMs) [8-10]. Recently, there has been a lot of interest in transition metal nitrides due to their distinctive mechanical, electronic, catalytic, optical, and magnetic characteristics for technological applications as well as their basic significance in condensed matter physics [11-19]. Various transition metal nitrides and their allotropes have also been reported and have attracted significant attention due to their unique physical and mechanical properties including high catalytic activity, chemical inertness, extreme hardness etc. [20-24].

Due to the extended lives and increased wear resistance that higher hardness offers, industries frequently seek superhard materials for tools utilized in heavy duty operations. These materials often need high-pressure, high-temperature (HPHT) conditions to form, thus, conventional ultra-hard materials like diamond and cubic boron nitride (*c*-BN) are expensive and challenging to make. Due to these restrictions, research has been concentrated on finding innovative superhard materials that are less expensive to manufacture. Compounds containing the light elements C, B, O, and N are shown to be effective candidates. The next generation of superhard materials is being paved by the transition metal borides, which among these contenders have great hardness and are easily produced under ambient circumstances [25]. It has been demonstrated through experiments that the N-to-W ratio in the WN compounds may vary, which correlates to various $W_xN_y$ crystals. As a result, the synthesized WN compounds have very complex structures. Early measurements showed the structures of a number of WN compounds, including hexagonal WN, [26] hexagonal and rhombohedral $W_2N_3$, cubic $W_3N_4$, [23] NaCl type WN [27]. Recently, Balasubramanian *et al.* [28] used the first-principles technique to explore the stability of WN compounds and discovered that whereas WN is mechanically unstable in cubic structures, it is mechanically stable in the NbO structure. It is important to note that because tungsten nitride exists in a number of phases with a wide range of compositions, it is likely that certain stable WN crystal forms were not visible in earlier investigations. The $WN_6$ is a structure that has just been predicted, and it is thermodynamically stable at high pressure [20]. Recently, first-principles simulations by Mounet *et al.* [29] demonstrate that two-dimensional $W_2N_3$ may easily exfoliate from a layered hexagonal bulk $W_2N_3$; the latter was first observed experimentally by Wang *et al.* in 2012 [23]. $W_2N_3$ is an excellent candidate for mechanical exfoliation because its binding energy is 26.3 meV, which is very similar to the values calculated for the most prevalent transition-metal dichalcogenides [23]. David Campi *et al.* [30] found intrinsic superconductivity in monolayer $W_2N_3$ with a critical temperature of 21 K, just above that of liquid hydrogen by means of first-principles calculations. Jing-Yang You *et al.* [31] reported a detailed investigation of the superconductivity and non-trivial electronic topology in 2D monolayer $W_2N_3$. They found that monolayer $W_2N_3$ is a superconductor with transition temperature of about 22 K and has a superconducting gap of 5 meV, based on the anisotropic Midgal-Eliashberg formalism.

To the best of our knowledge, only a few of the physical properties of bulk $W_2N_3$ have been studied so far, including its structural properties, bulk modulus and its pressure derivative, electronic properties (band structure, density of states and Fermi surface), phonon spectra and superconducting transition temperature [30, 32]. There are still many unexplored significant physical aspects of $W_2N_3$. Remarkably, most of the physical characteristics relevant to prospective applications, e.g., electronic charge density distribution, various mechanical properties, Mulliken bond population analysis, theoretical hardness, optical properties, thermodynamic properties of $W_2N_3$ have not been explored at all till date. The aim of this study is to look at these unexplored physical properties in detail. Some of the physical properties have been revisited for validation and completeness. The results presented herein shows that $W_2N_3$ is a highly promising compound for applications in the engineering and optoelectronic device sectors.

The remaining parts of this manuscript are structured as follows: In Section 2, an in-depth description of the computational methodology used in the present study can be found. In Section 3, an extensive discussion of the investigated properties and their possible consequences was presented. In Section 4, the important features from our study are summarized and discussed.

**2. Computational methodology**
The main DFT tool used in this investigation was the CASTEP code [33]. The potential for electronic exchange correlation was assessed using the Perdew–Burke–Ernzerhof scheme for solids (PBEsol) functional within the generalized gradient approximation (GGA) [34]. The Vanderbilt-type ultra-soft pseudopotential [35] was used to simulate the interactions between electrons and the ion cores. The basis sets for the valence electron states for W and N were $5s^25p^65d^46s^2$ and $2s^22p^3$, respectively. The first Brillouin zone in the reciprocal space of the hexagonal unit cell of $W_2N_3$ is integrated over using the Monkhorst-Pack (MP) technique [36] with a k-point mesh of 26×26×5 grid. The eigenfunctions of the valence and nearly valence electrons were expanded using a plane-wave basis at a cutoff energy of 550 eV. Using the Broyden-Fletcher-Goldfarb-Shanno (BFGS) technique [37], the internal forces and total energy were both relaxed during the geometry optimization. The highest ionic Hellmann-Feynman force was less than 0.03 eV/Å, the maximum ionic displacement was less than $1×10^{-3}$ Å, the maximum stress was less than 0.05 GPa and the tolerance for total energy was less than $10^{-5}$ eV/atom in order to accomplish the self-consistent convergence.

The DFT-based finite strain method [38] is used to determine the elastic characteristics. This approach relaxes the atomic degrees of freedom by applying a series of finite uniform deformations on the typical unit cell. The single crystal elastic constants $C_{ij}$ are then determined from the resulting stresses using a series of linear expressions:

$$\sigma_{ij} = \sum_{ij} C_{ij}\delta_{ij} \qquad (1)$$

where $\delta_{ij}$ stands for the six stress components that are applied to each strain on the conventional unit cell. The bulk and shear elastic moduli of polycrystalline materials, which are determined using this method, are calculated using the well-known Voigt-Reuss-Hill approximations [39–41]. The maximum force within 0.006 eV/Å and the maximum ionic displacement within $2×10^{-4}$ Å are fixed as the convergence condition for estimating the elastic characteristics.

The calculation of electronic charge density distribution and Fermi surface requires that k-point spacing has to be less than $0.01Å^{-1}$. So, we have chosen a k-point mesh of 40×40×7 grids in this work. Using the density functional perturbation theory (DPFT) based finite-displacement method (FDM) [42-43], which is embedded into the CASTEP code, the dynamical stability and lattice dynamic properties such as phonon dispersion and phonon density of states were calculated.

From the complex dielectric function $\varepsilon(\omega) = \varepsilon_1(\omega) + i\varepsilon_2(\omega)$, the frequency dependent optical constants of $W_2N_3$ have been derived. The momentum matrix elements between the unoccupied and the occupied electronic orbitals can be used to obtain the imaginary part of the dielectric function, $\varepsilon_2(\omega)$ by using the following equation:

$$\varepsilon_2(\omega) = \frac{2\pi e^2}{\Omega \varepsilon_0} \sum_{k,v,c} |\psi_k^c| \mathbf{u} \cdot \mathbf{r} |\psi_k^v|^2 \delta(E_k^c - E_k^v - E). \qquad (2)$$

where, $\Omega$ is the unit cell volume, $\omega$ is the angular frequency of the incident phonon, $e$ is the charge of an electron, $\psi_k^c$ and $\psi_k^v$ are respective wave functions for conduction band and valence band electrons at a specific $k$, and $\mathbf{u}$ is the unit vector defining the polarization direction of the incident electric field. Using the Kramers-Kronig transformation equation, the real part of the dielectric function, $\varepsilon_1(\omega)$, has been determined from the corresponding imaginary part, $\varepsilon_2(\omega)$. The other optical functions, namely, the absorption coefficient $\alpha(\omega)$, reflectivity $R(\omega)$, refractive index $n(\omega)$, energy loss-function $L(\omega)$, and optical conductivity $\sigma(\omega)$, can be deduced from the estimated values of $\varepsilon_1(\omega)$ and $\varepsilon_2(\omega)$ using the expressions found in the literature [44].

The quasi-harmonic Debye model, as implemented in the Gibbs program [45], is used to investigate the thermodynamic properties at different temperatures and pressures. We have utilized the *E-V* data and used the third-order Birch-Murnaghan equation of state (EOS) [46, 47] and equilibrium values of $E_0$, $V_0$, and $B_0$ obtained using the DFT method at zero temperature and zero pressure for finite temperature/pressure computations.

## 3. Results and discussion
*3.1 Structural optimization and phase stability of $W_2N_3$*

The optimized crystal structure of $W_2N_3$ is depicted in Fig. 1. $W_2N_3$ crystallizes in hexagonal structure with space group *P6₃/mmc* (No. 194) [23] and has a layered structure. There are two formula units and ten atoms per unit cell. Fully relaxed structure for $W_2N_3$ is obtained by optimizing the geometry including the lattice constants and internal atomic positions. The optimized W atom is located at the 4*f* Wyckoff position, with fractional coordinates (1/3, 2/3, 0.840), and N atom is located at the 4*f* and 2*c* Wyckoff positions, with fractional coordinate (1/3, 2/3, 0.070) and (1/3, 2/3, 1/4), respectively [23]. Table 1 represents the values of lattice constants *a* and *c*, equilibrium unit cell volume *V*, internal atomic coordinate *z*, and the formation enthalpy $\Delta H$ for $W_2N_3$ at ambient pressure. The lattice parameter of $W_2N_3$ was first reported in 2012 with *a* = 2.890 Å and *c* = 15.286 Å [23]. Y. Wang *et al.* reported the lattice parameters of $W_2N_3$ for the second time with *a* = 2.870Å and *c* = 15.175Å [48]. The optimized lattice parameters of our study are found to be 2.888 Å and 15.807 Å for *a* and *c*, respectively. These values are highly consistent with those found in previous studies. The negative value of enthalpy reveals that, at ambient pressure, tungsten nitride in the hexagonal structure is thermodynamically stable.

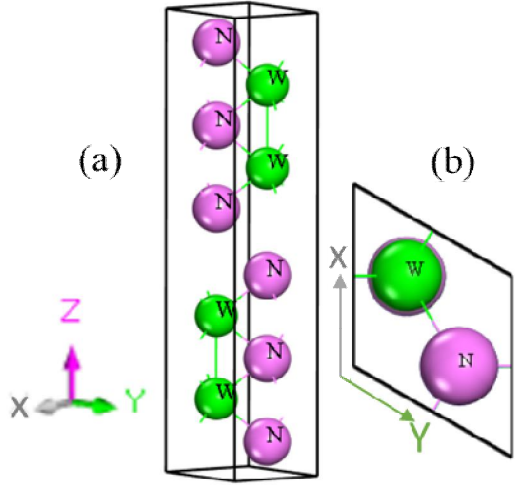

**Fig. 1.** (a) Conventional unit cell of $W_2N_3$ and (b) its 2D view in the xy-plane

**Table 1.** Calculated and previously obtained experimental/theoretical lattice constants ($a$, $b$, and $c$) (all in Å), unit cell volume ($V$ in Å$^3$), internal coordinate (z), and enthalpy of formation ($\Delta H$ in eV/atom) of $W_2N_3$.

| $a$ | $b$ | $c$ | $V$ | $z$ | $\Delta H$ | Ref. |
|---|---|---|---|---|---|---|
| 2.888 | 2.888 | 15.807 | 114.14 | 0.088 | -2.23 | This work |
| 2.890 | 2.890 | 15.286 | -- | --- | --- | [23][expt.] |
| 2.870 | 2.870 | 15.175 | 108.22 | --- | --- | [48][theo.] |

*3.2 Mechanical properties*

Elastic constants are very important material characteristics. The elastic constants of crystalline solids provide the link between mechanical and dynamical behavior of crystal under external stress of different types concerning the nature of the forces operating in solids, especially for the stability and stiffness of materials. The elastic constants are correlated with a material's mechanical characteristics, including stability, stiffness, brittleness, ductility, and elastic anisotropy. For engineering purposes, these characteristics are crucial when choosing a material for a specific task. According to the Born-Huang conditions, a hexagonal system has to satisfy the following inequality requirements in order to be mechanically stable [49]:

$$C_{11}-|C_{12}|> 0,$$
$$(C_{11}+C_{12})C_{33}-2C_{13}^2>0,$$

$$C_{44} > 0$$

$W_2N_3$ has positive values for each of its elastic constants $C_{ij}$ as shown in Table 2 and satisfies the above stability requirements, indicating that it is mechanically stable.

**Table 2.** Calculated elastic constants, ($C_{ij}$ in GPa), tetragonal shear modulus, ($C'$ in GPa) and the internal strain parameter ($\xi$) of $W_2N_3$.

| Compound | $C_{11}$ | $C_{12}$ | $C_{13}$ | $C_{33}$ | $C_{44}$ | $C_{66}$ | $C'$ | $\xi$ | Ref. |
|---|---|---|---|---|---|---|---|---|---|
| $W_2N_3$ | 535.1 | 204.0 | 7.7 | 38.8 | 11.8 | 165.5 | 165.5 | 0.522 | This work |
|  | 570 | 205.0 | 14.0 | 85.0 | 30.0 | -- | -- | -- | [48]$^{theo.}$ |

Every elastic constant has a different meaning; for example, the resistance to linear compressions in the [100] and [001] directions can be measured by the elastic constants $C_{11}$ and $C_{33}$, respectively. The bonding strength in $W_2N_3$ is stronger and compressibility is lesser along [100] direction than along [001] direction as $C_{11}$ is much greater than $C_{33}$. Elastic constant $C_{44}$ stands for the compound's resistance to shear deformation with respect to a tangential stress applied to the (100) plane in the [010] direction. According to our computed values, $C_{44}$ is substantially lower than $C_{11}$ and $C_{33}$. According to this $W_2N_3$ deforms more readily under shear than under unidirectional stress. The off-diagonal shear components are represented by the elastic constants $C_{12}$ and $C_{13}$, which are connected to the resistance of a compound as a result of shears in different crystal planes. The resistance of the (100) plane to shear in the [110] direction is correlated with the elastic constant $C_{66}$. $C_{44}$ has a somewhat lower value for the compound under study than $C_{66}$. For $W_2N_3$, $(C_{11}+C_{12}) > C_{33}$, which predicts that the elastic tensile modulus, is greater in the (001) plane than it is along the c-axis and thus the bonding in the (001) plane is elastically stiffer than that along the c-axis. Another measure of crystal's stiffness, known as the tetragonal shear modulus, is determined by the following equation:

$$C' = \frac{(C_{11} - C_{12})}{2} \qquad (3)$$

This parameter indicates the dynamic stability of a material. If $C'$ is greater than zero, the compound is stable; otherwise, it is dynamically unstable. In Table 2, the tetragonal shear modulus for $W_2N_3$ is reported to be 165.5 GPa (positive), thus $W_2N_3$ is predicted to be dynamically stable. A dimensionless internal strain parameter, known as the Kleinman parameter ($\xi$), is a measure of stability of a compound against stretching and bending. To compute this parameter for $W_2N_3$, the following equation has been used [50]:

$$\xi = \frac{(C_{11} + 8C_{12})}{(7C_{11} + 2C_{12})} \quad (4)$$

$\xi$ has the value in the range $0 \leq \xi \leq 1$. The upper and lower limits of $\xi$ are respectively represented by the significant contribution of bond bending to resist the external stress and the significant contribution of bond stretching to resist the external stress. The estimated values of $\xi$ of $W_2N_3$ is 0.522, from which it can be predicted that mechanical strength in $W_2N_3$ is almost equally affected by bond bending and bond stretching/contracting. Additionally, the Kleinman parameter describes how the cation and anion sub-lattices' relative position shifts in the crystal when volume-conserving distortions cause atomic locations to alter in ways not ensured by the ground state crystal symmetry.

While the elastic moduli produced via the Voigt-Reuss-Hill (VRH) approximation are often based on polycrystalline aggregates of compounds, the elastic constants estimated from DFT are based on single crystals. The following relationships [51-53] are used to estimate the Hill approximations [41] for the bulk modulus ($B_H$), shear modulus ($G_H$), Young's modulus ($Y$), Poisson's ratio ($v$), and hardness ($H$) of $W_2N_3$:

$$B_H = \frac{(B_V + B_R)}{2} \quad (5)$$

$$G_H = \frac{(G_V + G_R)}{2} \quad (6)$$

$$Y = \frac{9BG}{(3B + G)} \quad (7)$$

$$v = \frac{(3B - 2G)}{2(3B + G)} \quad (8)$$

$$H = \frac{(1 - 2v)}{6(1 + v)} \quad (9)$$

The elastic moduli (bulk modulus $B$, shear modulus $G$, and Young's modulus $Y$) characterize how the materials in the polycrystalline aggregates behave mechanically under loading. The resulting values are given in Table 3. The bulk modulus, $B$, shows resistance to fracture, whereas the shear modulus $G$, represents resistance to plastic deformation. For $W_2N_3$, a lower value of *G* relative to *B* as shown in Table 3 indicates that the mechanical strength will be constrained by plastic deformation. The bulk modulus (*B*) is inversely proportional to the cell volume (*V*) [54]. As a result, the bulk modulus *B*, which has a significant association with the cohesive energy or binding energy of the atoms in crystals, might be employed as a measure of the average atomic bond strength of materials [55]. On the other side, a high shear modulus value indicates that strong directional bondings are present between atoms [56]. Young's modulus can be used to calculate the resistance of a material to tension or compression along its length. The critical thermal shock coefficient is inversely proportional to the Young's modulus *Y* [57], which means that the Young's modulus has an impact on a material's ability to resist thermal shock. Better thermal shock resistance is associated with higher *R* values. For the selection of thermal barrier coating (TBC) materials, the thermal shock resistance is a crucial factor. The high value of *Y* as shown in Table 3 indicates that $W_2N_3$ is capable of high resistance to thermal shock. In general, larger values of Young's modulus suggest higher stiffness for a particular class of materials [58]. The calculated Young's modulus of our studied compound is quite medium; therefore, it is a moderately stiff material. Various thermophysical parameters and elastic moduli are interrelated. For example, a material's lattice thermal conductivity ($k_l$) and Young's modulus (*Y*) are connected as $k_l \sim \sqrt{Y}$ [59].

The factors, such as the Pugh's ratio (*G/B*) and the Poisson's ratio ($\sigma$) can characterize materials brittle or ductile nature [60-62]. The shear modulus to bulk modulus ratio (*G/B*) was suggested by Pugh [61] in 1954 as a useful measure for determining the brittleness and ductility of materials. 0.57 is the key value that distinguishes brittle from ductile materials. A value higher than 0.57 is associated with brittleness, whereas a value lower than 0.57 is correlated with ductility. Frantsevich *et al*. [62] also distinguished between brittleness and ductility in terms of the Poisson's ratio, and they proposed that 0.26 act as the boundary between brittle and ductile nature. The material will be brittle if the Poisson's ratio is less than 0.26; or else, the material will be ductile. The Pugh's ratio (*G/B*) and Poisson's ratio of $W_2N_3$ is 0.577 and 0.258, respectively. These values indicate that the mechanical state of $W_2N_3$ is located quite close to the brittle/ductile borderline.

The lower and upper limits of Poisson's ratio for a solid for central-forces are 0.25 and 0.50, respectively [63, 64]. We can predict from the values of Poisson's ratio as shown in Table 3 that the interatomic force of $W_2N_3$ is central in nature. The Poisson's ratio can be employed as well to identify if a material has covalent or ionic bonds. The values of $\sigma$ for ionic and covalent materials are generally 0.25 and 0.10, respectively [65]. Our computed Poisson's ratio of $W_2N_3$ is 0.258 suggesting that $W_2N_3$ contains ionic contribution.

The term *machinability* refers to the characteristic of a material that determines how easily it may be machined using a cutting tool. In engineering manufacturing and production, this parameter is used frequently. The work material, cutting tool, and cutting settings are a few of the variables that affect machinability. The choice of cutting tool material, tool shape, cutting force, feed rate, and depth of cut are all determined by the machinability of a given material. Furthermore, it determines the solid's dry lubricating properties and plasticity [66-69]. The formula for calculating a material's machinability index, $\mu_M$, is [70]:

$$\mu_M = \frac{B}{C_{44}} \tag{10}$$

A compound with a low $C_{44}$ value provides superior dry lubricity, according to this equation. A compound with a higher $B/C_{44}$ value has better lubricating qualities, lower feed forces, lower friction, and greater plastic strain values. $W_2N_3$ has a $B/C_{44}$ value of 8.62. This implies a very high level of machinability.

For the purpose of ensuring product quality in industry, the study of material hardness is of utmost importance. It illustrates how a material is affected by high loads. To understand elastic and plastic properties of a compound, measurement of hardness is crucial. The hardness of solid materials can be expressed into two broad categories. These are (i) soft material whose hardness is less than 10 GPa and (ii) hard material whose hardness is higher than 10 GPa [71-73]. $W_2N_3$ has estimated hardness of 9.69 GPa, which is reported in Table 3. This value indicates that $W_2N_3$ is moderately hard.

**Table 3.** The calculated bulk modulus ($B$ in GPa), shear modulus ($G$ in GPa), Young's modulus ($Y$ in GPa), Pugh's indicator ($G/B$), machinability index ($\mu_M$), Poisson's ratio ($\sigma$) and Vickers hardness ($H_V$ in GPa) of $W_2N_3$.

| Compound | B | G | Y | G/B | $\sigma$ | $\mu_M$ | $H_V$ | Ref. |
|---|---|---|---|---|---|---|---|---|
| $W_2N_3$ | 104.15 | 60.10 | 151.21 | 0.577 | 0.258 | 8.62 | 9.69 | This work |
|  | 131.00 | 84.00 | -- | 0.640 | -- | -- | 13.00 | [48][theo.] |

*3.3 Elastic anisotropy*

One important aspect that affects mechanical stability and structural strains of a material under various forms of stress is the elastic anisotropy. Anisotropy indices express the direction dependence mechanical characteristics of a system. Elastic anisotropy regulates a variety of physical processes, including the growth of plastic deformation in crystals, the propagation of microcracks in solids, the alignment/misalignment of quantum dots, phonon conductivity, and defect mobility. It also regulates the mechanical toughness of materials. Thus, an adequate understanding of anisotropic mechanical behavior is essential. Anisotropy and isotropy in crystals are typically dominated by covalent (directional) and metallic bonding, respectively [74, 75].

The degree of anisotropy in atomic bonding in various crystal planes can be determined by using the shear anisotropy factors. Three different factors given below can be used to measure the shear anisotropy in a hexagonal crystal [56, 76]:

$$A_1 = \frac{4C_{44}}{(C_{11} + C_{33} - 2C_{13})} \tag{11}$$

$$A_2 = \frac{4C_{55}}{(C_{22} + C_{33} - 2C_{23})} \tag{12}$$

$$A_3 = \frac{4C_{66}}{(C_{11} + C_{22} - 2C_{12})} \tag{13}$$

where $A_1$, $A_2$ and $A_3$ are the respective shear anisotropy factors for {100}, {010}, {001} planes between <011> and <010> directions, <101> and <001> directions, and <110> and <010> directions, respectively.

Since $C_{11} = C_{22}$, $C_{44} = C_{55}$ and $C_{13} = C_{23}$ for hexagonal crystals, thus $A_1 = A_2$. When $A_1 = A_2 = A_3$, then the crystal is said to be isotropic with respect to shear, otherwise it is anisotropic. In Table 4, the estimated values of shear anisotropy of $W_2N_3$ are listed.

The following standard equations can be used to determine the universal anisotropy index ($A^U$), equivalent Zener anisotropy measure ($A^{eq}$), anisotropy in shear ($A_G$), and anisotropy in compressibility ($A_B$) of materials with any crystal symmetry.

$$A^U = 5\frac{G_V}{G_R} + \frac{B_V}{B_R} - 6 \geq 0 \tag{14}$$

$$A^{eq} = \left(1 + \frac{5}{12}A^U\right) + \sqrt{\left(1 + \frac{5}{12}A^U\right)^2} \tag{15}$$

$$A^G = \frac{(G_V - G_R)}{2G^H} \tag{16}$$

$$A^B = \frac{(B_V - B_R)}{(B_V + B_R)} \tag{17}$$

One of the most used indices for measuring anisotropy in elastic characteristics is the universal anisotropy index ($A^U$). Regardless of the crystal symmetry, it is a single measure of anisotropy. Contrary to all other known anisotropy measures, $A^U$ is the anisotropy parameter which takes account of both shear and bulk contributions. We may infer from Eqn. (14), that $G_V/G_R$ has a stronger impact on the anisotropy index $A^U$ than $B_V/B_R$. The value of $A^U$ less than zero or larger than zero indicates variable degrees of anisotropy, whereas $A^U$ is zero for an isotropic material. The estimated $A^U$ of W$_2$N$_3$ is 19.82, which demonstrates highly anisotropic characteristics. For locally isotropic crystals, $A^{eq}$ equals 1.0. At ambient pressure the estimated values of $A^{eq}$ for W$_2$N$_3$ is 18.47, which also indicates that the crystal is highly anisotropic. $A^G$ and $A^B$ have values between 0 and 1. The ideal elastic isotropy and the maximum elastic anisotropy are represented, respectively, by $A^G = A^B = 0$ and $A^G = A^B = 1$. The values of $A^G$ and $A^B$ are respectively 0.62 and 0.65 which are listed in Table 4. These values imply that W$_2$N$_3$ has slightly larger anisotropy in compressibility than in shear.

The universal log-Euclidean anisotropy index is defined by a log-Euclidean formula as follows [77, 78]:

$$A^L = \sqrt{\left[\ln\left(\frac{B_V}{B_R}\right)\right]^2 + 5\left[\ln\left(\frac{C_{44}^V}{C_{44}^R}\right)\right]^2} \tag{18}$$

where $C_{44}^V$ and $C_{44}^R$ are, respectively, the estimated values of $C_{44}$ from the Voigt and the Reuss limits. These values can be obtained as [77]:

$$C_{44}^V = \frac{5}{3}\frac{C_{44}(C_{11} - C_{12})}{3(C_{11} - C_{12}) + 4C_{44}} \tag{19}$$

and

$$C_{44}^R = C_{44}^V + \frac{3}{5}\frac{(C_{11} - C_{12} - 2C_{44})^2}{3(C_{11} - C_{12}) + 4C_{44}} \tag{20}$$

The expression for $A^L$ is valid for every crystal symmetry, same as the universal anisotropy index. This index is true for all crystallographic point symmetry groups and is scaled appropriately for perfect isotropy. But when investigating extremely anisotropic crystallites, $A^L$ is shown to be less sparse than $A^U$, making it more relevant for the present study. The absolute amount of anisotropy cannot be explained by $A^U$; only the anisotropic nature. Hence, the difference between the averaged stiffness of $C^V$ and $C^R$ is used in $A^L$ calculations, and it is thought to be more suitable for anisotropy studies. The range of $A^L$ values is 0 to 10.26. Nearly 90% of solids have $A^L$ values greater than 1. $A^L$ is 0 in the event of perfect isotropy. The calculated value of $A^L$ is 5.31, which is much greater than 1, suggesting high level of anisotropy. According to theory, higher (lower) $A^L$ value indicates the presence of layered (non-layered) type structure [79-81]. In this study, the high value of $A^L$ for $W_2N_3$ indicates strongly layered structure. Such layered feature can make a compound highly suitable for exfoliation and chemical intercalation.

Along *a*- and *c*-axis, the linear compressibility of a hexagonal crystal can be evaluated by using the following equations [82]:

$$\beta_a = \frac{(C_{33} - C_{13})}{D} \quad \text{and} \quad \beta_c = \frac{(C_{11} + C_{12} - 2C_{13})}{D} \tag{21}$$

with $D = (c_{11} + c_{12})c_{33} - 2(c_{13})^2$

The estimated values of $\beta_a$, $\beta_c$, and $\beta_c/\beta_a$ are listed in Table 4. Crystals that are elastically isotropic have a unit value of $\beta_c/\beta_a$. The degree of elastic anisotropy in compression is quantified by the deviation of these factors from their unit value. Our estimated values imply once again that $W_2N_3$ is highly anisotropic.

**Table 4.** Shear anisotropy factor ($A_1$, $A_2$ and $A_3$), the universal anisotropy index $A^U$, equivalent Zener anisotropy measure $A^{eq}$, anisotropy in shear $A_G$, anisotropy in compressibility $A_B$, universal log-Euclidean index $A^L$, linear compressibility ($\beta_a$ and $\beta_c$) (TPa$^{-1}$), and their ratio $\beta_c/\beta_a$ for $W_2N_3$.

| Phase | $A_1$ | $A_2$ | $A_3$ | $A^U$ | $A^{eq}$ | $A_G$ | $A_B$ | $A^L$ | layered | $\beta_a$ | $\beta_c$ | $\beta_c/\beta_a$ |
|---|---|---|---|---|---|---|---|---|---|---|---|---|
| $W_2N_3$ | 0.085 | 0.085 | 1.0 | 19.83 | 18.47 | 0.62 | 0.65 | 5.31 | Yes | 0.0012 | 0.0281 | 23.5 |

With the help of the ELATE program [83], the directional dependencies of the Young's modulus (*Y*), linear compressibility (*β*), shear modulus (*G*), and Poisson's ratio (*v*) of the $W_2N_3$ can be examined. Using the CASTEP code, the elastic stiffness matrices needed for this investigation are computed. The isotropic nature of crystals is manifested in the uniform circular 2D and spherical 3D graphical representations. The degree of anisotropy increases with the deviations from these ideal shapes. Figs. 2 show the 3D view of $Y$, $\beta$, $G$, and $\upsilon$ for $W_2N_3$ together with the 2D projection on the *xy*-, *zx*-, and *yz*-planes. The minimum and maximum values of the parameters are shown by the curves in green and blue colors, respectively. From two dimensional representations, it is clear that all of the four parameters are anisotropic in *zx*- and *yz*-planes but are isotropic in the *xy*-plane. From 3D plots, it can be observed that the 3D figures of $Y$, $\beta$, $G$, and $v$ show a large departure from spherical form, indicating the degree of anisotropy. The anisotropy order is displayed graphically in 2D and 3D plots as $\upsilon > G > \beta > Y$. In addition, ELATE describes a quantitative analysis that lists the maximum and minimum values of $Y$, $\beta$, $G$, $\upsilon$ and their ratios, as listed in Table 5.

**Table 5.** The minimum limit, maximum limit and anisotropy of Young's modulus ($Y$, in GPa), compressibility ($\beta$, in TPa$^{-1}$), shear modulus, ($G$, in GPa), and Poisson's ratio ($v$) of $W_2N_3$.

| Phase | Young's modulus | | | Linear compressibility | | | Shear modulus | | | Poisson's ratio | | |
|---|---|---|---|---|---|---|---|---|---|---|---|---|
| | $Y_{min}$ | $Y_{max}$ | $A_Y$ | $\beta_{min}$ | $\beta_{max}$ | $A_\beta$ | $G_{min}$ | $G_{max}$ | $A_G$ | $v_{min}$ | $v_{max}$ | $A_v$ |
| $W_2N_3$ | 32.78 | 456.78 | 13.93 | 1.09 | 25.33 | 23.23 | 11.84 | 165.57 | 13.98 | 0.0104 | 0.5804 | 56.03 |

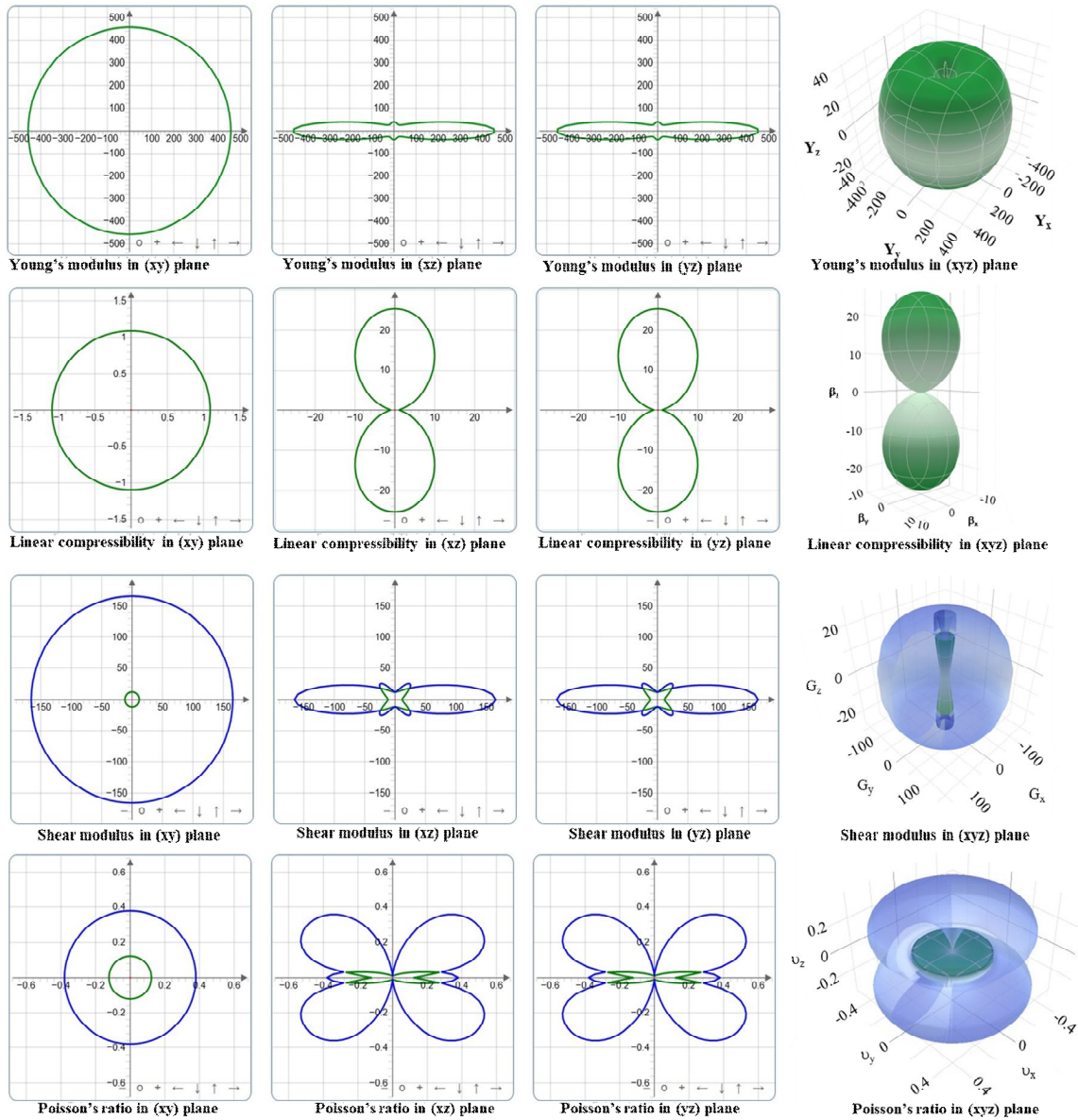

**Fig. 2.** Direction-dependence of Young's modulus ($Y$), compressibility ($\beta$), shear modulus ($G$), and Poisson's ratio ($v$) of $W_2N_3$ single crystal.

*3.4 Acoustic velocities and its anisotropy*

A material's sound velocity is a significant characteristic that is correlated with its electrical and thermal conductivity. In recent years, physics, materials science, the design of musical instruments, seismology, geology, and the medical sciences have all taken notable interest in investigating the acoustic behavior of compounds and composites. Thermal conductivity is increased in a crystal structure with a higher sound velocity ($v$). The following equations [84-86] show how the bulk and shear moduli are related to the speed of transverse and longitudinal sound waves traveling through a crystalline material:

$$v_l = \left(\frac{3B + 4G}{3\rho}\right)^{1/2} \quad \text{And} \quad v_t = \left(\frac{G}{\rho}\right)^{1/2} \tag{22}$$

Using the following equation [84], the average sound velocity in a polycrystalline system is calculated from the transverse and longitudinal sound velocities:

$$v_m = \left[\frac{1}{3}\left(\frac{2}{v_t^3} + \frac{1}{v_l^3}\right)\right]^{-1/3} \tag{23}$$

In Table 6, the calculated acoustic velocities have been tabulated.

Each atom in a solid has three vibrational modes: one longitudinal and two transverse. Pure longitudinal and transverse modes only occur along certain axes in an anisotropic crystal. The propagating modes, on the other hand, are either quasi-transverse or quasi-longitudinal in all other directions. In $W_2N_3$, pure transverse and longitudinal modes can only exist for the symmetry directions along [100] and [001] due to the hexagonal symmetry of the compound. The single crystal elastic constants can be used to calculate the acoustic velocities of $W_2N_3$ in these principal directions [87]:

[100] direction:

$$[100]_{v_l} = \sqrt{\frac{(C_{11}-C_{12})}{2\rho}}; \quad [010]_{v_{t1}} = \sqrt{C_{11}/\rho}; \quad [001]_{v_{t2}} = \sqrt{C_{44}/\rho} \tag{24}$$

[001] direction:

$$[001]_{v_l} = \sqrt{C_{33}/\rho}; \quad [100]_{v_{t1}} = [010]_{v_{t2}} = \sqrt{(C_{44}/\rho)} \tag{25}$$

where $v_l$ is the longitudinal sound velocity, $\rho$ is the crystal density, and $v_{t1}$ and $v_{t2}$ are the first and second transverse acoustic modes, respectively. Table 7 lists the computed sound velocities for these directions. The longitudinal velocity of $W_2N_3$ along [100] is significantly higher than that along [001].

**Table 6.** Density $\rho$ (g/cm$^3$), transverse sound velocity $v_t$ (ms$^{-1}$), longitudinal sound velocity $v_l$ (ms$^{-1}$), and average sound wave velocity $v_m$ (ms$^{-1}$) of $W_2N_3$.

| Phase | $\rho$ | $v_t$ | $v_l$ | $v_m$ | Ref. |
| --- | --- | --- | --- | --- | --- |
| $W_2N_3$ | 14.28 | 2051.50 | 3592.35 | 2279.71 | This work |

**Table 7.** Anisotropic sound velocities (in ms$^{-1}$) in W$_2$N$_3$ along principal crystallographic directions.

| Phase | Propagation directions | | Sound velocity |
|---|---|---|---|
| W$_2$N$_3$ | [100] | $[100]_{vl}$ | 3404.87 |
| | | $[010]_{vt1}$ | 6121.44 |
| | | $[001]_{vt2}$ | 909.03 |
| | [001] | $[100]_{vl}$ | 1648.36 |
| | | $[010]_{vt1}$ | 909.03 |
| | | $[001]_{vt2}$ | 909.03 |

*3.5 Phonon dispersion – phonon DOS and phonon dynamics*

The phonon dispersion spectra (PDS) and phonon density of states (PHDOS) can be employed to determine a variety of characteristics of a material. For example, about dynamical stability, phase transitions, and vibrational contributions of atoms to thermal expansion, heat capacity, and Helmholtz free energy from the phonon dispersion spectrum. The dynamical stability of a material is a crucial criterion for applications involving a time-varying applied loading. The phonon density of states and the electron-phonon interaction are intimately connected. The total phonon density of states in the ground state and the calculated phonon dispersion spectra of W$_2$N$_3$ in the high symmetry directions of the Brillouin zone (BZ) are shown in Fig. 3. If the phonon frequencies over the whole BZ are positive, a compound is expected to be dynamically stable. Soft phonon modes and dynamic instability are ensured to exist when negative phonon frequencies are present. Since W$_2$N$_3$ has 10 atoms per unit cell and the total number of phonon modes is three times the total number of atoms per unit cell, W$_2$N$_3$ contains 30 phonon modes. It has three acoustic modes colored by pink lines as shown in Fig. 3. There are 27 optical modes since a unit cell made up of *N* atoms has three acoustic modes and (3*N*-3) optical modes. The coherent oscillations of atoms in a lattice outside of their equilibrium position are what give rise to acoustic phonons. In contrast, when one atom moves to the left and its neighbor moves to the right, the lattice's atoms oscillate out of phase, giving rise to the optical phonon. At $\Gamma$ point, W$_2$N$_3$ exhibits the highest optical frequency and it is 20.80 THz.

The lattice dynamics of crystalline solids is particularly important for the zone-center phonon modes. Among the 27 optical modes, 12 are Raman active, 6 are IR active and the other 9 are silent modes. The irreducible representations of the Brillouin zone-center optical phonon modes can be categorized in accordance with the factor group theory [88] as follows:

$$\Gamma_{opt.} = 2A_{2u} + 4E_{1u} + 6E_{2g} + 4E_{1g} + 2A_{1g} \tag{26}$$

where, $A_{2u}$ and $E_{1u}$ are IR active and $E_{2g}$, $E_{1g}$ and $A_{1g}$ are Raman active and $B_{2g}$, $E_{2u}$ and $B_{1u}$ are silent modes. When two or more modes have the same frequency yet cannot be distinguished from one another, they are referred to as degenerate modes. This is why there are six IR active modes and twelve Raman active modes included in Table 8. The highest frequencies observed in the IR and Raman active modes are 20.08 THz and 20.12 THz, respectively.

To see the contribution of each band to various atomic modes of vibration, we have also calculated the total and atomic partial PHDOS for $W_2N_3$, which are displayed alongside the PDS. The PHDOS curve indicates that, whereas the higher optical branches (with frequencies > 15.7 THz) originate mainly from the vibration of lighter N-atoms, the acoustic and lower optical modes arise due to the vibration of heavier W atoms. Peaks in PHDOS are produced due to the flatness of the bands and the heights of the peaks in the total PHDOS are decreased due to the wide band dispersion. For W and N atoms, the prominent peaks in the PHDOS are seen around 5.27 THz and 20.10 THz, respectively.

**Table 8.** Theoretical wave-numbers $\omega_i$ and symmetry assignment of the IR-active and Raman-active modes of $W_2N_3$.

| Phase | Mode | | Irr. Rep. | Wave-numbers, $\omega$ (cm$^{-1}$) |
|---|---|---|---|---|
| $W_2N_3$ | IR | $\omega_1$ | $A_{2u}$ | 540.0 |
| | | $\omega_2$ | $A_{2u}$ | 669.3 |
| | | $\omega_3$ | $E_{1u}$ | 358.5 |
| | | $\omega_4$ | $E_{1u}$ | 476.1 |
| | Raman | $\omega_1$ | $E_{2g}$ | 19.4 |
| | | $\omega_2$ | $E_{2g}$ | 351.2 |
| | | $\omega_3$ | $E_{2g}$ | 476.4 |
| | | $\omega_4$ | $E_{1g}$ | 80.3 |
| | | $\omega_5$ | $E_{1g}$ | 409.9 |
| | | $\omega_6$ | $A_{1g}$ | 180.3 |
| | | $\omega_7$ | $A_{1g}$ | 670.6 |

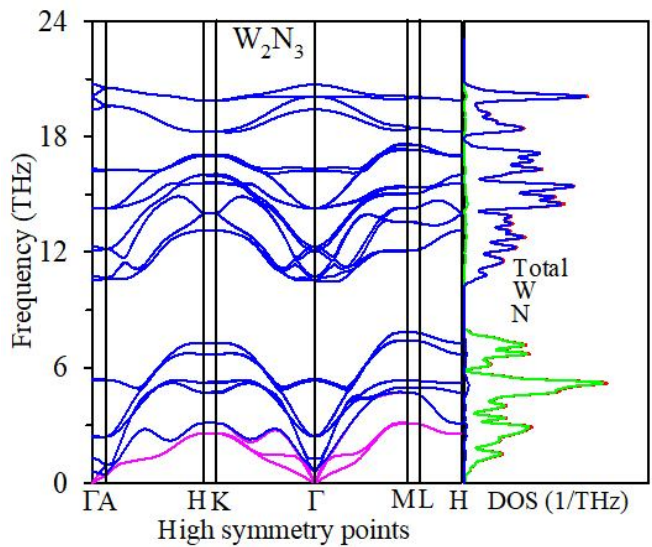

**Fig. 3.** The phonon dispersion spectra (PDS) and phonon density of states (PHDOS) of $W_2N_3$.

*3.6 Bonding character – charge density distribution*

A useful tool for determining the type of interatomic chemical bonding is the electronic charge distribution map. It demonstrates how electrical charges around various atomic species are accumulated or depleted. Covalent bonding between two atoms is demonstrated by the accumulation of charges between them. A negative and positive charge balance at the atom locations is utilized to predict the presence of ionic bonds. On the other hand, uniform charge smearing shows metallic bonding. The electronic charge density in various crystal planes is shown in Fig. 4 in order to understand the chemical bonding between the atoms of $W_2N_3$. The overall electron density is shown on the right-hand side of charge density maps using a color scale in the unit of $e/Å^3$, with blue and red denoting high and low charge (electron) densities, respectively. Fig. 4 clearly shows that W atoms have comparatively high electron density than N atoms; covalent bonding exists between W and N atoms. Accumulation of charge also exists between W-W and N-N atoms. But, in the latter case the degree of covalency is comparatively weaker than that of W-N bonding. Hence, the compound possesses a mixture of covalent and metallic bonding.

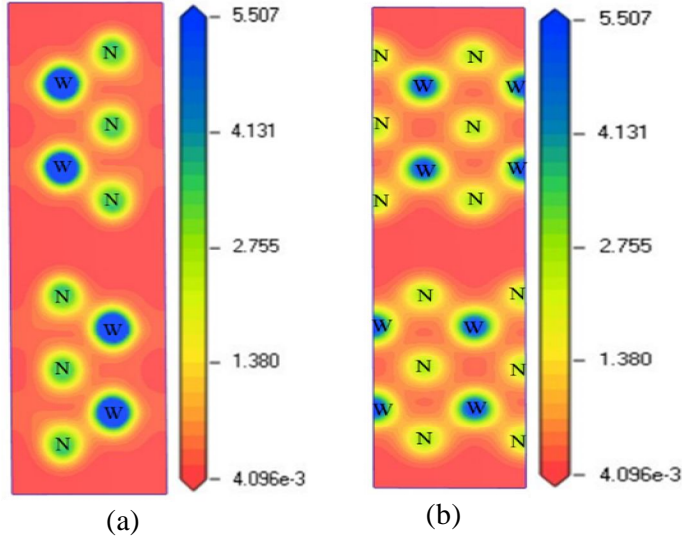

**Fig. 4.** Charge density distribution maps of $W_2N_3$ in (a) (111) and (b) (011) plane.

*3.7 Thermophysical parameters*

*3.7.1. Debye temperature*

The temperature at which the wavelength of phonons of a material nearly matches to the interatomic spacing is generally known as the Debye temperature $\varTheta_D$. It also defines a border line between the lattice vibration's classical and quantum characteristics. The high- and low-temperature behavior of a solid can be separated using this temperature. In order to comprehend several thermophysical features of solids, like melting temperature, bonding forces, thermal conductivity, the energy required for the creation of vacancies, specific heat, phonon dynamics, and superconductivity, it is crucial to study the Debye temperature. All vibrational modes have nearly the same energy, $k_B T$, when $T > \varTheta_D$. However, it is found that at $T < \varTheta_D$, the higher frequency modes are frozen. Debye temperature can be estimated using a variety of methods. When temperatures are low, only acoustic modes are responsible for the vibrational excitations. As a result, at low temperatures, the Debye temperature estimated from elastic constants agrees with that calculated from the specific heat measurement. In this study, the Debye temperature of $W_2N_3$ is calculated by Anderson technique, using the following equation [89]:

$$\varTheta_D = \frac{h}{k_B} \left[ \frac{3n}{4\pi V_0} \right]^{1/3} v_m \tag{27}$$

where $n$ is the number of atoms within a unit cell, $V_0$ is the volume of a unit cell, $k_B$ is Boltzmann's constant, $h$ is Planck's constant, and $v_m$ is mean sound velocity. Hexagonal $W_2N_3$ has a Debye temperature of ~380 K in the ground state which is enlisted in Table 9.

The Debye temperature of $W_2N_3$ is moderate which suggests that the atomic bonding strengths are not very strong and the material under study is not very hard in nature.

*3.7.2 Phonon thermal conductivity*

The transfer of heat through the vibrations of lattice ions within a solid is quantified by lattice (phonon) thermal conductivity. It is one of the most crucial thermal factors in determining the energy conversion efficiency of thermoelectric materials. Both phonons and electrons can carry thermal energy in solids. At low temperatures, electrons are the primary heat carriers in metals. The study of lattice thermal conductivity is essential for materials intended for high temperature applications. With a vast array of technical applications, including the development of novel thermoelectric materials, sensors, heat sinks, transducers, and thermal barrier coatings, the phonon thermal conductivity of solids, is one of the key thermophysical parameters. With a formula developed by Slack [90-92], the lattice thermal conductivity, $k_{ph}$ as a function of temperature can be estimated as follows:

$$k_{ph} = A(\gamma)\frac{M_{av}\theta_D^3\delta}{\gamma^2 n^{2/3} T} \tag{28}$$

where $\gamma$ is the Grüneisen parameter, $T$ is the absolute temperature, $n$ is the total no. of atoms in the unit cell, $M_{av}$ is the average atomic mass (in kg/mol) in a crystal, $\delta$ is the cubic root of the average atomic volume and $\theta_D$ is the Debye temperature $A(\gamma)$ is the $\gamma$ dependent parameter that can be calculated from the following equation [93]:

$$A(\gamma) = \frac{4.85628 \times 10^7}{2\left(1 - \frac{0.514}{\gamma} - \frac{0.228}{\gamma^2}\right)} \tag{29}$$

The room temperature (300 K) value of the calculated lattice thermal conductivity, $k_{ph}$, is enlisted in Table 9. Callaway–Debye theory [94] states that the lattice thermal conductivity at low temperatures is directly proportional to the Debye temperature, $\theta_D$. The lattice thermal conductivity of a material increases with increasing $\theta_D$. Furthermore, a material's lattice thermal conductivity and Young's modulus are correlated as: $K_{ph} \propto \sqrt{Y}$ [95]. The phonon thermal conductivity of $W_2N_3$ is high at room temperature (Table 9).

*3.7.3 Grüneisen parameter*

An important thermophysical parameter which estimates the anharmonic effects in a solid is known as the Grüneisen parameter $\gamma$. It is associated with several significant physical processes, including thermal conductivity, thermal expansion, acoustic wave absorption, and the temperature dependence of elastic characteristics. The larger the value of $\gamma$, the higher is degree of anharmonicity. Using Poisson's ratio, the Grüneisen parameter of $W_2N_3$ can be estimated by using the following equation [96]:

$$\gamma = \frac{3(1+v)}{2(2-3v)} \qquad (30)$$

The estimated value of $W_2N_3$ is 1.54, which is shown in Table 9. This value is typical for solids.

*3.7.4 Melting temperature*

A parameter of interest that restricts the temperature range in which a solid can be applied is the melting temperature ($T_m$). A solid will exhibit stronger atomic interaction, higher bonding energy, higher cohesive energy and lower coefficient of thermal expansion if it has high $T_m$ [85, 97-99]. Solids can be continually utilized below $T_m$ without chemical change or excessive distortion producing mechanical difficulties. With the use of elastic constants, the following empirical relationship can be used to determine a material's melting point [98]:

$$T_m = 345K + (4.5K/GPa)\left(\frac{2C_{11}+C_{33}}{3}\right) \qquad (31)$$

The estimated melting temperature of $W_2N_3$ is 2008.5 K, which is listed in Table 9. Thus, $W_2N_3$ can be used as a promising candidate material for high temperature applications due to its high melting temperature. High heat of fusion, low fusion entropy, or a combination of both is the main causes of a high melting point.

*3.7.5 Heat capacity*

In addition to being necessary for many applications, the specific heat capacity offers crucial information into the material's vibrational characteristics. The constant volume heat capacity, $C_v$, goes to the Dulong-Petit limit at high temperatures and it is proportional to $T^3$ at very low temperature [99]. The specific heat capacity, at constant-volume, $C_v$, and at constant-pressure, $C_p$, for $W_2N_3$ with different temperatures at $P = 0$ GPa and different pressures at $T = 300$ K are respectively shown in Figs. 5 (a, b) and (c, d). The heat capacities of $W_2N_3$ increase as temperature rises due to phonon thermal softening, as seen in Fig. 5 (a, c). The specific heat capacities, $C_v$ and $C_p$, exhibit a significant rise up to around 300 K as a result of the anharmonic approximation of the Debye model. As seen in Fig. 5 (c), at high temperature ($T > 300K$), the anharmonic impact on $C_v$ is suppressed for $W_2N_3$, and $C_v$ approaches the Dulong-Petit limit, which is typical for all solids. With the increase of pressure, $C_v$ and $C_p$ are seen to decrease as shown in Fig. 5 (b, d). The Debye temperature obtained within the quasi-harmonic approximation is ~399 K, quite close to the value estimated using the sound velocities.

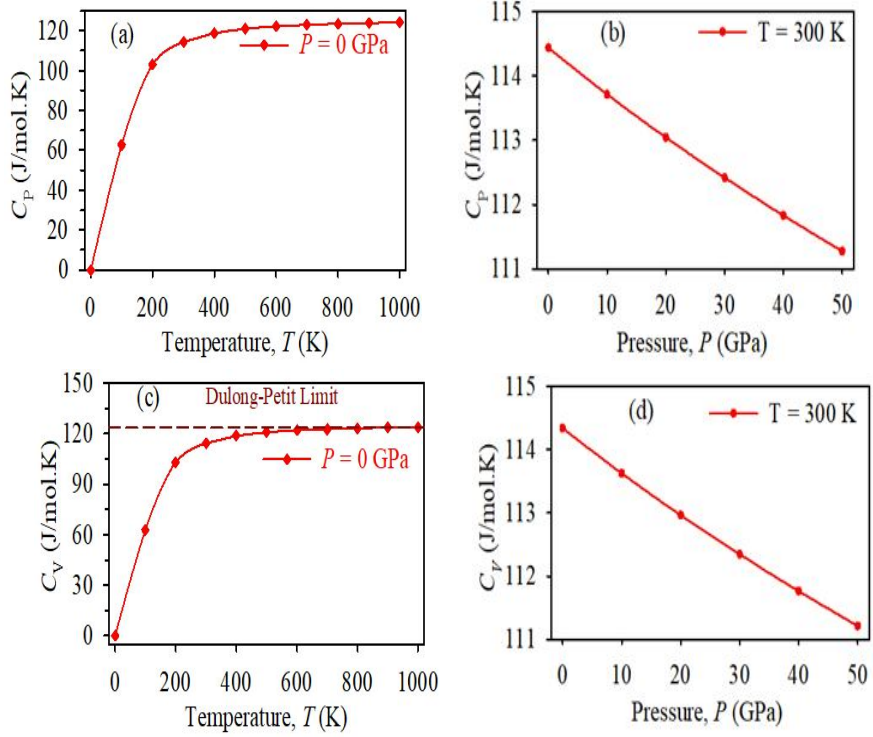

**Fig. 5.** Temperature and pressure dependent variations of specific heat capacities $C_p$ and $C_v$ of $W_2N_3$.

*3.7.6 Entropy*

The entropy, $S$, is a broad aspect of a thermodynamic system that measures the amount of disordered energy content in a material. Figure 6 (a and b) demonstrate the change in entropy, $S$, as a function of temperature and pressure. With the rise of temperature, the entropy is seen to increase [Fig. 6(a)] as a consequence of increasing thermal disorder. Additionally, for $T = 300$ K, it is seen from Fig. 6(b) that the entropy falls with the increase of pressure. This is because when pressure increases, the volume of the solid drops.

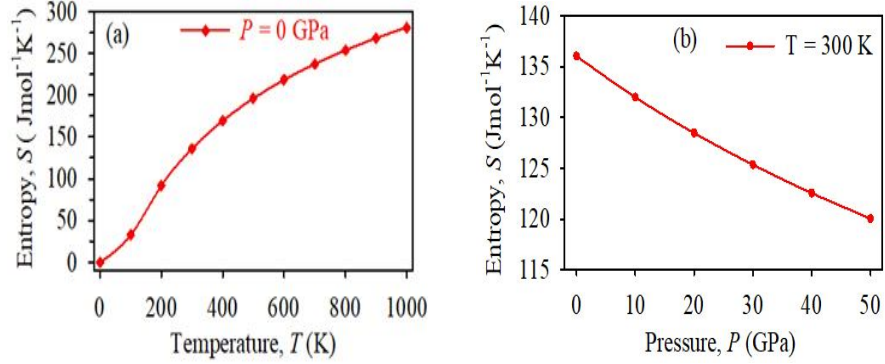

**Fig. 6.** Temperature and pressure dependent variations of entropy, $S$, of $W_2N_3$.

**Table. 9.** Calculated melting temperature ($T_m$ in K), Debye temperature ($\Theta_D$ in K), lattice thermal conductivity, $k_{ph}$ (W/m-K) at 300 K, and Grüneisen parameter, $\gamma$ for $W_2N_3$.

| Compound | $T_m$ | $\Theta_D$ | | $k_{ph}$ | $\gamma$ | Ref. |
|---|---|---|---|---|---|---|
| | | Using elastic constant | Using QHD model | | | |
| $W_2N_3$ | 2008.5 | 380.2 | 399.3 | 16.5 | 1.54 | This |

*3.8 Electronic properties*

*3.8.1 Band structure*

In order to understand the electronic, optical, and magnetic characteristics of materials at the microscopic level, it is important to understand their electronic band structure. The effective masses of charge carriers can be calculated from band structure. It also greatly influences the charge transport and bonding properties. The nature of dominating bands close to the Fermi level can be used to better understand a material's charge transport characteristics. The electronic energy band structure of $W_2N_3$ has been calculated and is depicted in Fig. 7 along different high symmetry directions ($\Gamma$-$A$-$H$-$K$-$\Gamma$-$M$-$L$-$H$) in the momentum space. The horizontal broken line at zero energy indicates the Fermi level ($E_F$). The unit cell of $W_2N_3$ has 84 different energy bands in all. It is evident from Fig. 7 that there is no band gap at the Fermi level. This demonstrates the metallic character of hexagonal $W_2N_3$ in the optimized structure. The bands that cross the Fermi level are displayed in various colors along with the band numbers that correlate to each color. Mainly N-*2p* and W-*5d* states contribute to the energy bands near the Fermi level. This indicates that N-*2p* and W-*5d* states dominate the charge transport properties of $W_2N_3$. It is noteworthy that the band crossing close to the $\Gamma$-point exhibits hole-like characteristic. All the bands crossing the Fermi level are found to be fairly dispersive. We can better understand the underlying Fermi surfaces by using the band structure calculations.

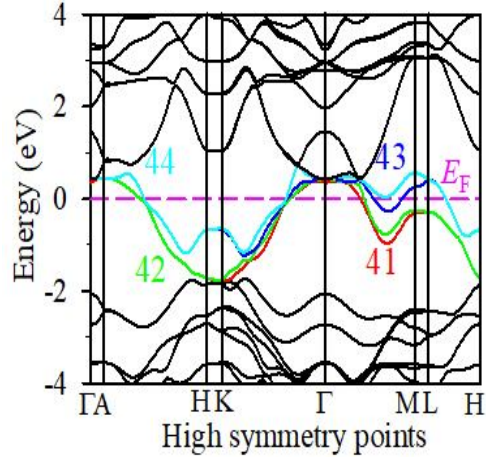

**Fig. 7.** Electronic band structure of $W_2N_3$ along several high symmetry directions in the Brillouin zone.

*3.8.2 Density of states*

The number of electronic states that are accessible to be occupied at each energy level per unit energy interval is referred to as the electronic energy density of states, or simply DOS. The structure of the DOS in the valence and conduction bands is connected to almost all of the electrical and optical characteristics of crystalline solids. The contribution of different atoms or orbitals to optoelectronic properties of a material can be understood by studying its total and partial density of states. The DOS of a material is also crucial to understand the contribution of each atom to bonding and antibonding states. Fig. 8 depicts the calculated total density of states (TDOS) and atom resolved partial density of states (PDOS) of $W_2N_3$. The Fermi level, $E_F$, is shown by the vertical broken line at 0 eV. $W_2N_3$ exhibits metallic electrical conductivity, as indicated by the non-zero value of TDOS at the Fermi level $E_F$. It is found that $W_2N_3$ has a TDOS value of ~4.0 states per eV per unit cell or ~2.0 states per eV per formula unit at $E_F$. The highly dispersive bands crossing the Fermi level gives this low value of $N(E_F)$. In order to explain how W and N atoms contribute to TDOS and chemical bonding, the PDOS of these atoms has also been determined. In the vicinity of $E_F$, W *5d* and N *2p* dominantly contribute to the TDOS of $W_2N_3$. Their respective values are 1.65 and 2.16 states per eV per unit cell at the Fermi level. The TDOS has large peaks at -3.77 and 2.88 eV that are close to the $E_F$. These bonding or anti-bonding peaks are the results of the hybridizations of W *5d* and N *2p* electronic orbitals. Such hybridization close to the Fermi energy is frequently taken as an indication of the creation of strong covalent bonds.

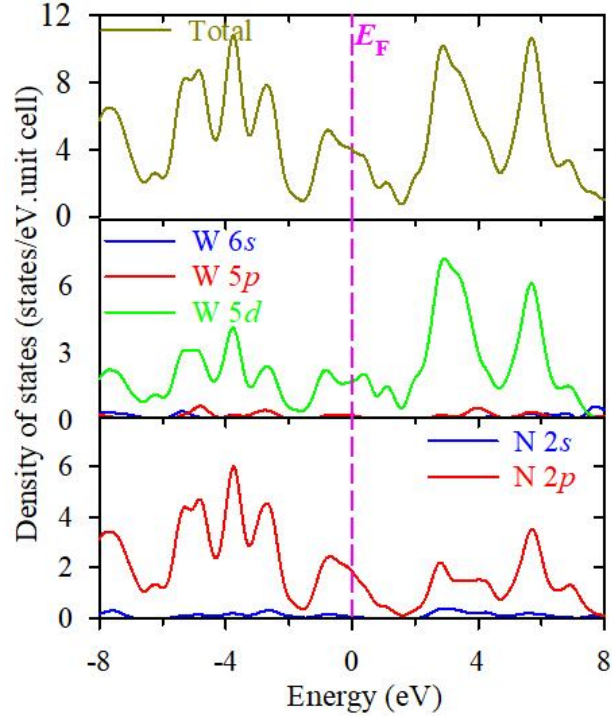

**Fig. 8.** Total and partial electronic density of states of $W_2N_3$. The vertical line shows the Fermi energy.

*3.8.3 Coulomb pseudopotential*

The Coulomb pseudopotential is the measure of itinerant electron-electron interaction in a metal. It explains how coulomb repulsion affects superconductivity. The coulomb pseudopotential can be estimated by using the following equation [100]:

$$\mu^* = \frac{0.26 N(E_F)}{1 + N(E_F)} \tag{32}$$

The calculated value of the Coulomb pseudopotential is 0.173. This suggests that the electronic correlation in $W_2N_3$ is significant. This mainly arises due to significant contribution of the W *5d* electrons to the TDOS at the Fermi level. The effective electron-phonon interaction that causes the Cooper pairs to develop in the context of superconductivity is reduced by the Coulomb pseudopotential. The superconducting transition temperature, $T_c$, decreases as a consequence [100-103].

*3.8.4 Fermi surface*

The Fermi surface (FS) topology controls the electronic properties of metals. The Fermi surface separates the occupied electronic states from the unoccupied electronic states at low temperature. The topology of a Fermi surface has a significant impact on a number of characteristics, including electronic, optical, thermal, and magnetic ones. The Fermi surface topology of $W_2N_3$ is shown in Fig. 9. (a, b, c and d). The 41, 42, 43 and 44 numbered bands cross the Fermi level (shown in Fig. 7), and are responsible for the construction of Fermi surface. The Fermi surface of $W_2N_3$ is made up of four Fermi sheets of different geometry. The FSs for the bands 41 and 42 are quite similar. Both the structures have circular-like sheets close to the center of the BZ. These are 2D electron-like sheets. On the other hand, the FSs for 43 and 44 bands are also similar and comparatively complex in shape. Here prismatic-like hexagonal cross sections are seen around the *G*-A direction. The remaining of these topologies consists of six separate parts parallel to the L-M directions. These are hole-like and located at the corner of the BZ. This implies that both electron- and hole-like behaviors exist in $W_2N_3$. The Fermi surface topology also indicates that electronic transport is anisotropic in $W_2N_3$. Close resemblance between the Fermi surfaces for bands 41 and 42, and bands 43 and 44 show that the pair of bands are highly degenerate.

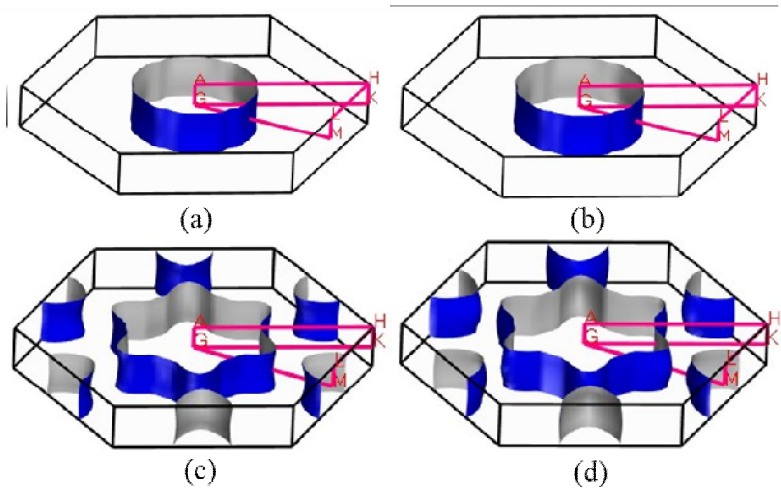

**Fig. 9.** Fermi surfaces of $W_2N_3$ for the band number (a) 41, (b) 42, (c) 43, and (d) 44.

*3.9 Optical properties*

The optical properties of a material indicate how it reacts to incident electromagnetic radiation. From the perspective of optoelectronic applications, the response to visible light is crucial. In many areas of modern science and technology, including display devices, sensors, lasers, photo-electrodes, photo-detectors, photonics, solar cells, etc., the study of optical properties of solids has attracted significant interest. Additionally, optical anisotropy must be taken into account since various common optical technologies, such as 3D movie screens, LCD displays, polarizers, and wave plates, are developed using this [104]. Various energy/frequency dependent optical properties, notably the dielectric function, loss function, refractive index, optical conductivity, absorption coefficient, and reflectivity can completely determine the response of a material to incident light. We have estimated the optical properties of $W_2N_3$ for photon energies up to 30 eV for [100] and [001] polarization directions of the electric field as shown in Fig. 10 (a-h) to investigate the possible anisotropy. Electronic band structure and energy dependent density of states characteristics govern optical properties. For the investigation of optical properties, it is necessary to include Drude damping for metallic compound [105-108]. Since the band structure and density of states of $W_2N_3$ show that it is metal, a Drude damping of 0.05 eV and plasma frequency of 5 eV are used in order to calculate its optical properties.

The variations of the real and imaginary parts of the dielectric function, $\varepsilon(\omega)$, with respect to the photon energy, are respectively shown in Fig. 10 (a) and (b). In Fig. 10 (a), the real part, $\varepsilon_1(\omega)$, of the dielectric function vanishes at around 27 eV, which corresponds to a peak in the energy loss function as shown in Fig. 10 (h). The metallic conductivity of $W_2N_3$ is shown by the fact that $\varepsilon_1(\omega)$ goes through zero from below (negative value). It is also clear from Fig. 10 (b) that $W_2N_3$ is metallic in nature since the imaginary part of the dielectric function in both directions approaches to zero from above. Fig. 10 (b), also shows that the values of $\varepsilon_2$ drop to zero at 28 eV, suggesting that the material will become transparent above 28 eV energy.

A dimensionless quantity that characterizes how light travels through a medium is known as the refractive index. It is closely related to the local field inside the material as well as the electronic polarizability of ions. For designing photoelectric devices, the complex refractive index is a very significant factor. The phase velocity of the electromagnetic wave inside the sample is determined by the real part of the refractive index, whereas the extinction coefficient (imaginary part) spectrum reveals how much the incident electromagnetic radiation is attenuated while passing through the material. The frequency dependence of the refractive index of $W_2N_3$ for [100] and [001] polarization directions are respectively shown in Figs. 10 (c) and (d). For both directions, refractive index of $W_2N_3$ is large at low energy and decreases with increasing energy. Due to high static refractive index value of $W_2N_3$, it can be used in display devices.

The absorption coefficient ($\alpha$) is the measure of the ability of a material to absorb incoming electromagnetic radiation. It is a useful measure for determining a material's electronic properties, regardless of whether it is metallic, semiconducting, or insulating [109]. The absorption coefficient plays a very important role to know how well a semiconducting material converts solar energy and how much light of specific energy can enter the material before being absorbed. The energy dependent absorption spectra of $W_2N_3$ for the polarization directions [100] and [001] are illustrated in Fig. 10 (e). In this figure, the nonzero value at zero photon energy is due to the metallic character of $W_2N_3$, which is compatible with the dielectric function, DOS, and band structure calculations. For the [100] and [001] polarizations, the highest absorption is seen at about 9.6 eV and 19.1 eV, respectively. Significant optical anisotropy can be observed in the absorption properties in this way. From Fig. 10 (e), it is seen that, $\alpha$ decreases sharply at ~ 26.5 eV for both polarization directions, which agrees well with the position of loss peak as shown in Fig. 10 (h).

The conduction of free charge carriers over a certain range of photon energy can be characterized from the optical conductivity of a material. This is a dynamic response of mobile charge carriers, including the electron-hole pairs produced by photons in semiconductors. The real part of the photoconductivity ($\sigma$) spectra of $W_2N_3$ is shown in Fig. 10 (f). At zero photon energy, the nonzero photoconductivity for both directions manifests that, $W_2N_3$ has no electronic band gap, which agrees with previously calculated band structure and density of states. For $W_2N_3$, the maximum photoconductivity is obtained at zero photon energy for both polarizations. $W_2N_3$ shows isotropic nature at low energy region and anisotropic nature after 1.26 eV. Generally, the low energy (infrared) portion of the spectra is dominated by the intraband contribution to the optical characteristics. The interband transition, on the other hand, causes peaks in the high energy region of the absorption and conductivity spectra. $W_2N_3$ exhibits the highest peaks in the visible and ultraviolet regions.

The reflectivity spectra of $W_2N_3$ along [100] and [001] polarizations as a function of incident photon energy is shown in Fig. 10 (g). The reflectivity of $W_2N_3$ exhibits significant optical anisotropy. From Fig. 10 (g), it is seen that, at ambient pressure the reflectivity of $W_2N_3$ begins from zero frequency with a value of 0.99. The reflectivity remains above 90% in the infrared region. Fig 10 (g) shows that $W_2N_3$ has lower reflectivity in the entire visible light regions in addition to the low-energy part of the UV region of the solar spectra. Nevertheless, $W_2N_3$ has above 44% reflectivity in the visible region and can be employed as a good solar heat reflector [107].

An important optical parameter that describes how much energy a fast moving electron loses when moving through a material is known as the loss function. In the dielectric formalism used to explain the optical spectra and the excitations created by fast charges in solids, the energy loss function of a material is a crucial quantity. A material's absorption, reflection and loss function properties are interconnected. The plasma resonance is connected with the peaks in the loss function spectrum, and the frequency that corresponds to those peaks is known as the plasma frequency ($\omega_p$) [107]. The energy/frequency dependent electron energy loss function for $W_2N_3$ is depicted in Fig. 10 (h). It is interesting to note that the maxima of the loss function of $W_2N_3$ are found at 26.75 eV and 27.50 eV for [100] and [001] polarization directions, respectively. The sudden decrease in absorption coefficient and reflectivity of $W_2N_3$ as shown in Figs. 10 (e) and (g) can be linked to these sharp loss peaks. These frequencies (energies) are called bulk screened plasma frequency as the peak in loss function is associated with the plasma resonance. When $\varepsilon_2 < 1$ and $\varepsilon_1 = 0$, then the energy loss peak manifests in the high energy region [110, 111]. $W_2N_3$ will be transparent, and will switch from metallic to dielectric response if the incoming light frequency is greater than the plasma frequency.

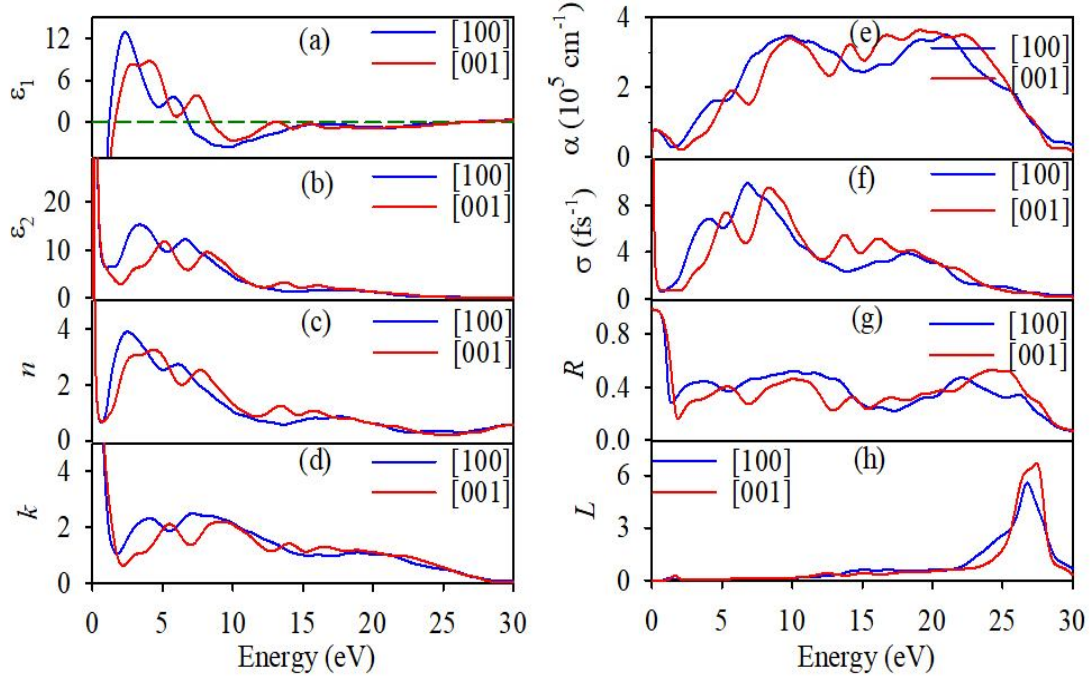

**Fig. 10.** Optical parameters of $W_2N_3$ as a function of photon energy.

*3.10 Thermodynamic properties*

The thermodynamic properties of $W_2N_3$ are evaluated in the temperature range of 0 to 1000 K and pressure range of 0 to 50 GPa pressure using quasi-harmonic approximation. The bulk modulus of a material is used to calculate its resistance to uniform compression. It also provides details on how well the material bonds together. Figs.11 (a) and 11 (b) reveal the temperature and pressure dependence of the isothermal bulk modulus of $W_2N_3$. According to our findings, at temperatures below 150 K, the bulk modulus of $W_2N_3$ is almost flat; at temperatures over 150 K, it drops gradually as shown in Fig.11 (a). From Fig. 11 (b) it is seen that the bulk modulus of $W_2N_3$ increases with increasing pressure, which satisfies the general formula for bulk modulus, $B = v\frac{\Delta p}{\Delta v}$.

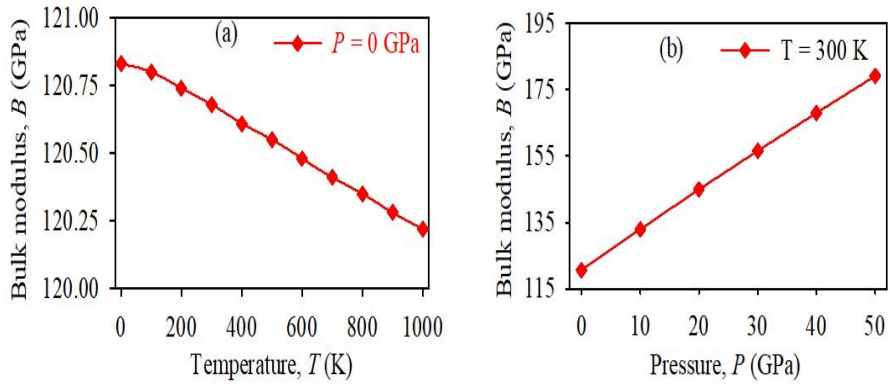

**Fig. 11.** Temperature and pressure dependent bulk modulus of $W_2N_3$.

Figs. 12 (a) and (b) illustrate the volume thermal expansion coefficient (VTEC) with respect to temperature and pressure, respectively. Up to 300 K, the coefficients of $W_2N_3$ under investigation grow quickly; however, the increment is slow above 300 K. Conversely, as pressure is increased, the expansion coefficient falls but at different rates at a constant temperature of 300 K. It has been demonstrated that there is an inverse relationship between the bulk modulus and volume thermal expansion coefficient of a material.

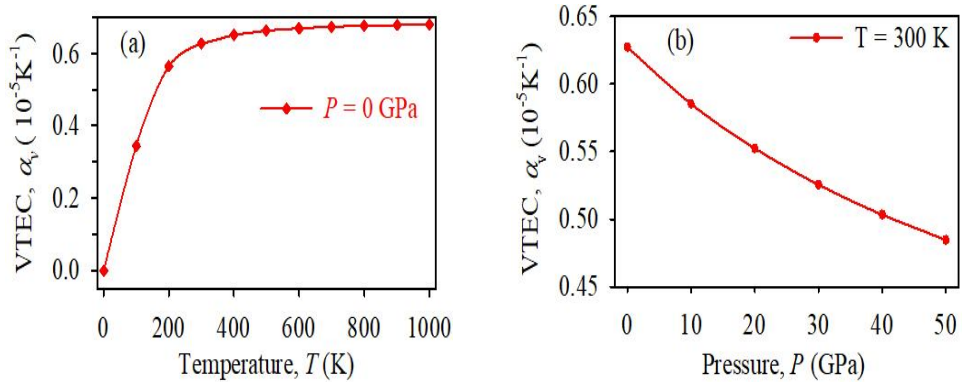

**Fig. 12.** Temperature and pressure dependent variations of volume thermal expansion coefficient of $W_2N_3$.

The internal energy, denoted as $U$, is the energy content of a material due to all the activated degrees of freedom within. Figs. 13 (a and b), respectively, illustrates how the internal energy of $W_2N_3$ varies with temperature and pressure. The internal energy of $W_2N_3$ thus grows almost linearly with temperature above 100 K. Fig. 13(b) shows that the internal energy of $W_2N_3$ rises also almost linearly as pressure rises.

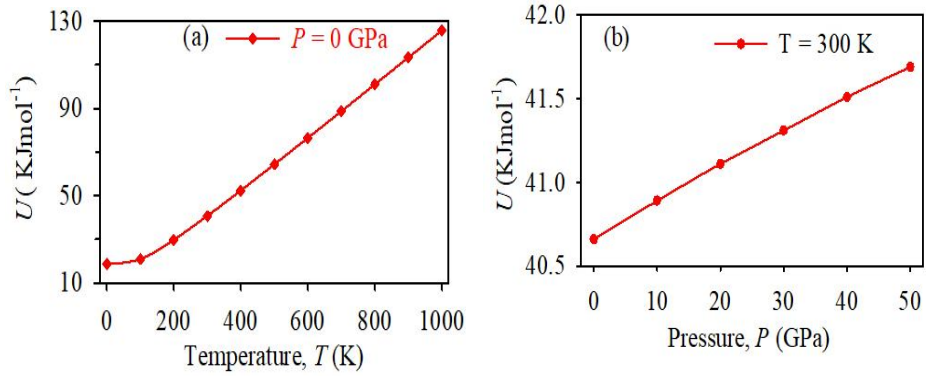

**Fig. 13.** Temperature and pressure dependent variations of internal energy of $W_2N_3$.

## 4. Conclusions

The unexplored mechanical, vibrational, elastic, thermophysical, electronic, optical, and thermodynamic characteristics of $W_2N_3$ have all been investigated in this paper. Our investigation demonstrates the thermodynamical and mechanical stability of $W_2N_3$. Phonon dispersion curves represent an illustration of the dynamical stability of our compound. $W_2N_3$ has a highly layered crystal structure, and is elastically anisotropic. $W_2N_3$ has borderline brittle characteristics, and its machinability level is extremely high. $W_2N_3$ has medium hardness. Combination of these features makes $W_2N_3$ an attractive compound for machine-tools device sector like some other engineering materials including the MAX and MAB phases [112-116]. The mechanical strength in $W_2N_3$ is controlled by both bond bending and bond stretching contributions. The charge density distribution of $W_2N_3$ shows direction dependence. There are significant covalent and metallic bondings. The substance being studied has a high melting point and a rather high phonon thermal conductivity at ambient temperature. Thus, it has potential to be used as a heat sink material. The electronic band structure shows metallic behavior with significant electronic correlations. The Fermi surface contains hole-like segments. The temperature and pressure dependent thermodynamic properties of $W_2N_3$ are conventional. The optical properties are anisotropic. The compound is an excellent infrared light reflector, and has a strong absorptivity of ultraviolet light. The reflectivity remains above 44% in the entire visible region, and it has high low-energy refractive index. These features make $W_2N_3$ suitable for optical device applications.

In conclusion, $W_2N_3$ has a number of appealing mechanical, thermal, and optoelectronic properties that make it a good fit for use in engineering and optical device applications. It is our hope that the new findings in this work will encourage future theoretical and experimental investigations on $W_2N_3$.

## Acknowledgements

S. H. N. and R. S. I. acknowledge the research grant (1151/5/52/RU/Science-07/19-20) from the Faculty of Science, University of Rajshahi, Bangladesh, which partly supported this work. I. A. acknowledges the fellowship from the Ministry of Science and Technology, Bangladesh, for his M.Phil. research.

## Data availability

The data sets generated and/or analyzed in this study are available from the corresponding author on reasonable request.

## CRediT author statement

**Istiak Ahmed:** Methodology, Software, Formal analysis, Writing-Original draft. **F. Parvin:** Supervision, Writing-Reviewing and Editing. **R. S. Islam:** Writing-Reviewing and Editing. **S. H. Naqib:** Conceptualization, Supervision, Formal analysis, Writing- Reviewing and Editing.

## Competing Interests

The authors declare no competing interests.